\documentclass[onecolumn,showpacs,preprintnumbers,amssymb,pra]{revtex4}
\usepackage{graphicx}
\usepackage{dcolumn}
\usepackage{color}
\usepackage{bm}

\begin{document}

\title{\bf Teleportation via maximally and non-maximally entangled mixed states}
\author{Satyabrata Adhikari}
\altaffiliation{tapisatya@gmail.com}
\affiliation{Dept. of Mathematics, Birla Institute of Technology, Mesra, Ranchi-835215,
India}
\author{Archan S. Majumdar}
\altaffiliation{archan@bose.res.in}
\affiliation{Dept. of Astro Phys. and Cosmology, S. N. Bose National Centre of Basic Sciences, Salt lake,
Kolkata-700098, India}
\author{Sovik Roy}
\altaffiliation{sovik1891@gmail.com}
\affiliation{Dept. of Mathematics, Techno India, EM 4 / 1 - Salt Lake City, Kolkata-700091, India}
\author{Biplab Ghosh}
\altaffiliation{quantumroshni@gmail.com}
\affiliation{Dept. of Phys., Vivekananda College for Women, Barisha,
Kolkata 700008, India}
\author{Nilkantha Nayak}
\affiliation{Centre for Applied Physics, Central Univ. of Jharkhand,
835205, India}

\date{\today}

\begin{abstract}
We study the efficiency of two-qubit mixed entangled states as resources
for quantum teleportation. We first consider two maximally entangled mixed
states,  viz., the Werner state\cite{werner}, and a class of states
introduced by Munro {\it et al.} \cite{munro}. We show that the Werner
state when used as teleportation channel, gives rise to better average
teleportation fidelity compared to the latter class of states for any
finite value of mixedness. We then introduce a non-maximally entangled
mixed state obtained as a convex combination of a two-qubit entangled
mixed state and a two-qubit separable mixed state. It is shown that such
a teleportation channel can
outperform another non-maximally entangled channel, viz., the Werner
derivative for a certain range of mixedness. Further, there exists a range of
parameter values where the former state satisfies a Bell-CHSH
type inequality and still performs better as a teleportation
channel compared to the Werner derivative even though the latter violates
the inequality.
\end{abstract}


\maketitle

\section{Introduction}

Quantum teleportation is one of the most relevant applications of quantum
information processing. Teleportation requires the
separation of a protocol into classical and quantum parts using
which it is possible to reconstruct an unknown input state
with perfect fidelity
at another location while destroying the original copy.
The original idea of teleportation
introduced by Bennett {\it et al.} \cite{bennett} is implemented
through a channel involving a pair of particles in a Bell
State shared by the sender and the receiver. Later, Popescu
\cite{popescu} showed that  pairs in a mixed state could
be used for imperfect teleportation. Further, it has been
shown that if the two distant parties adopt a "measure-and-prepare"
strategy for teleporting an unknown quantum state, then the
average fidelity of teleportation  is at most
$2/3$ which is the maximum fidelity achievable by means of
local operations and classical communications  \cite{popescu,massar,gisin}.
A quantum channel could be useful for communication purposes only if
its teleportation fidelity exceeds $2/3$.

In practice it is difficult to prepare pure states, but rather the states
obtained are generally mixed in their characteristics. Naturally, a
question arises as to whether better average teleportation fidelities compared
to that in classical protocols could be obtained  if mixed states were used in
quantum communication purposes. Therefore, the basic objective is to look
for such mixed states which when used as quantum teleportation
channels, give fidelity of teleportation higher than the classical
fidelity $2/3$. It has been found  that
Werner states \cite{werner} used as quantum teleportation
channels give higher teleportation fidelity\cite{horodecki1}. Recently,
the mixed state obtained from the Buzek-Hillery
cloning machine\cite{buzek-hillery} as a
teleportation channel has been studied\cite{adhikari}.

Similar to the case of pure entangled states, entangled mixed states
can also be
divided into two categories: (i) maximally entangled mixed states (MEMS)
and (ii) non-maximally entangled mixed states (NMEMS). Those states
that achieve the greatest possible entanglement for a given
mixedness are known as MEMS, otherwise they are NMEMS. The notion of MEMS
was first introduced by Ishizaka and Hiroshima \cite{ishizaka}.
They proposed a class of bipartite mixed states and
showed that entanglement of those states cannot be increased
further by any unitary operations (e.g., the Werner state). Later,
Munro et.al. (MJWK)\cite{munro} studied a class of states which has the
maximum amount of entanglement for a given degree of purity and
derived an analytical form for that class of MEMS. Apart from
maximally entangled mixed states, there are also NMEMS which can
be studied for some particular interest. Hiroshima and Ishizaka
\cite{hiroshima} studied a NMEMS called Werner derivative which
can be obtained by applying a unitary transformation on the Werner state.

The motivation for this work lies in performing a comparitive study of
mixed states in their capacity to perform as efficient channels for
quantum teleportation. It is known that not all entangled mixed states
are useful for teleportation \cite{horodecki}. So the question arises as
to whether all MEMS states could be used as teleportation channels. To this
end we first explore the capability of the MJWK class
of states \cite{munro} as teleportation channels by finding their
average teleportation fidelity. We find an upper bound for mixedness beyond
which the MJKW class of states is not useful for teleportation. We further
show that Werner states always act
as better teleportation channels for all finite values of mixedness,
even though they are less entangled compared to MJWK states for
a given entropy.
We then focus on non-maximally entangled mixed states and probe a question:
is there any family of NMEMS which outperforms existing NMEMS such as the
Werner derivative states\cite{hiroshima} when
used for quantum communication purposes ? To address this issue
we construct a new
entangled mixed state which is the convex combination of an entangled
mixed state and a separable mixed state. Our state is NMEMS since it
does not fall in the class of Ishizaka and Hiroshima's \cite{ishizaka}
MEMS. We show that this class of NMEMS can serve better as quantum
channel for teleportation compared to the Werner derivative for a range of
values of mixedness.

The relation between nonlocality of states
as manifested by the violation of Bell-CHSH inequalities\cite{chsh}
and their ability to perform as efficient teleportation channels is
interesting. It has been shown that there exist mixed states that do not
violate any Bell-CHSH inequality, but still can be used for
teleportation\cite{popescu}. Here we raise this question first with regard to
MEMS states and show there exists states in this category which satisfy the
Bell-CHSH inequality, but could be still useful for teleportation. We
then consider NMEMS states and find a range of parameters
for which our constructed state satisfies a Bell-CHSH type inequality
but still outperforms the Werner derivative in teleportation, even though
the latter violates the Bell-CHSH inequality. Finally, our comparitive
study of teleportation by maximally and non-maximally entangled mixed states
reveals that whereas in the former case, one class of states, i.e., Werner
states, definitely outperforms another, i.e., MJWK states for all
values of mixedness, the result for the NMEMS states that we consider
depends on their degree of mixedness.

The paper is organized as follows. In
section-II, we recapitulate some useful definitions and general
results related to mixed states, their violation of local inequalities,
and the optimal teleportation fidelities when they are used as
teleportation channels. We illustrate these general results with the
well-known example of the Werner state \cite{kim}. In section-III, we
study the efficiency of the MJWK states \cite{munro} in teleportation.
We then consider two different NMEMS in Section-IV.
We first study the Werner derivative\cite{hiroshima} as a teleportation
channel and also obtain the range of parameter values for which
it violates the
Bell-CHSH inequality. We next
introduce another NMEMS and investigate its entanglement properties
and efficiency as a teleportation channel. We further show
that this new NMEMS satisfies the Bell-CHSH
inequality. In Section-V we present a comparitive analysis of the MEMS
as well as the NMEMS channels for teleportation, also highlighting
their respective status vis-a-vis the Bell-CHSH inequality. Finally,
we summarize our results in Section-VI.

\section{The Werner state as a teleportation channel}

The Werner state is a convex combination
of a pure maximally entangled state and a maximally mixed state.
Ishizaka and Hiroshima \cite{ishizaka} showed that the entanglement of formation
\cite{wootters} of the Werner state cannot be increased by any unitary
transformation. Therefore, the Werner state can be
regarded as a maximally entangled mixed state. In this section we will
review the performance of the Werner state as a teleportation channel.
Though most of the results presented here are well known \cite{kim}, our
discussion is intended to set the stage for the analysis of other MEMS
and NMEMS states that we perform later. To begin with, let us recall certain
useful definitions on the entanglement, teleportation capacity and mixedness
of general states.

The maximal singlet fraction is defined
for a general state $\rho$ as \cite{bose}
\begin{eqnarray}
F(\rho)= {\mathrm max} \langle\Psi|\rho|\Psi\rangle
\label{sing.frac.}
\end{eqnarray}
where the maximum is taken over all maximally entangled states
$|\Psi\rangle$.

The linear entropy  $S_L$ for a mixed state $\rho$ is
defined by \cite{munro}
\begin{eqnarray}
S_{L}=\frac{4}{3}(1-Tr(\rho^{2})) \label{lin.ent.def.}
\end{eqnarray}

The concurrence for a bipartite state $\rho_{AB}$
is defined as \cite{wootters}
\begin{eqnarray}
C= max \{0,\lambda_{1}-\lambda_{2}-\lambda_{3}-\lambda_{4}\}
\label{concurrence}
\end{eqnarray}
where $\lambda$'s are the square root of eigenvalues of
$\rho\tilde{\rho}$ in decreasing order. The spin-flipped density
matrix $\tilde{\rho}$ is defined as
\begin{eqnarray}
\tilde{\rho}=(\sigma_{y}^{A}\otimes\sigma_{y}^{B})\rho^{*}(\sigma_{y}^{A}\otimes\sigma_{y}^{B})
\label{spin-flip}
\end{eqnarray}

The efficiency of a quantum channel used for teleportation is measured in
terms of its average teleportation fidelity given by \cite{horodecki2}
\begin{eqnarray}
f_{opt}^T(\rho_{\phi}) = \int_S d M(\phi) \sum_k p_k Tr (\rho_k \rho_{\phi})
\label{telfid0}
\end{eqnarray}
where $\rho_{\phi}$ is the input pure state and $\rho_k$ is the output state
provided the outcome $k$ is obtained by Alice. The quantity
$Tr (\rho_k \rho_{\phi})$ which is a measure of how the resulting state is
similar to the input one, is averaged over the probabilities of outcomes $p_k$,
and then over all possible input states ($M$ denotes the uniform distribution
on the Bloch sphere $S$). It has been shown \cite{horodecki} that if a state
is useful for standard teleportation, the optimal teleportation fidelity
can be expressed as
\begin{eqnarray}
f_{opt}^{T}(\rho)=\frac{1}{2}[1+\frac{N(\rho)}{3}]
\label{tel.fid.N}
\end{eqnarray}
where $N(\rho)=\sum_{i=1}^{3}\sqrt{u_{i}}$ and $u_{i}$'s are the
eigenvalues of the matrix $T^{\dagger}T$. The elements of the
matrix $T$ are given by
\begin{eqnarray}
t_{nm}=Tr(\rho~\sigma_{n}\bigotimes\sigma_{m})
\label{tmatrix}
\end{eqnarray}
 where $\sigma_{i}$'s
denote the Pauli spin matrices. Now, in terms of the quantity $N(\rho)$,
a general result\cite{horodecki} holds that
any mixed spin-$\frac{1}{2}$ state is useful
for (standard) teleportation if and only if
\begin{eqnarray}
N(\rho)>1
\label{teleportcond}
\end{eqnarray}
The relation between the optimal teleportation
fidelity $f_{opt}^{T}(\rho)$ and the maximal singlet fraction
$F(\rho)$ is given by \cite{horodecki1}
\begin{eqnarray}
f_{opt}^{T}(\rho)= \frac{2F(\rho)+1}{3} \label{T18}
\end{eqnarray}
From Eqs.(\ref{tel.fid.N}) and (\ref{T18}) it follows that
\begin{eqnarray}
F(\rho)=\frac{1+N(\rho)}{4}
\label{singletnrho}
\end{eqnarray}
Now using the inequality \cite{verst}
\begin{eqnarray}
F\leq\frac{1+N}{2}\leq\frac{1+C}{2}
\label{singletnegativity}
\end{eqnarray}
where $N$ denotes the negativity of the state, we have
\begin{eqnarray}
N(\rho)\leq 1+2N
\label{ineq123}
\end{eqnarray}

We now recall a useful result on the the violation of the
Bell-CHSH inequality by  mixed states. Any state described by the 
density operator
$\rho$ violates the Bell-CHSH inequality \cite{chsh} if and only if the
inequality
\begin{eqnarray}
M(\rho)=max_{i>j}(u_{i}+u_{j})>1
\label{bellineq}
\end{eqnarray}
holds, where
$u_{i}$'s are eigenvalues of the matrix $T^{\dagger}T$ \cite{horodecki}.

Let us now review the Werner state as a resource for teleportation \cite{kim}.
Though the
Werner state can be represented in various ways,  in the present work
we express it in terms of the maximal singlet fraction.
The Werner state can be written in the form
\begin{eqnarray}
\rho_{W}&&=
\frac{1-F_{w}}{3}I_{4}+\frac{4F_{w}-1}{3}|\Psi^{-}\rangle\langle\Psi^{-}|{}\nonumber\\&&=
\left(\begin{matrix}{\frac{1-F_{w}}{3} & 0 & 0 & 0\cr 0 &
\frac{1+2F_{w}}{6}& \frac{1-4F_{w}}{6} &0 \cr 0 &
\frac{1-4F_{w}}{6} & \frac{1+2F_{w}}{6} & 0 \cr 0 &0 & 0 &
\frac{1-F_{w}}{3} }\end{matrix}\right)
\label{werner}
\end{eqnarray}
where $|\Psi^{-}\rangle=\frac{|01\rangle-|10\rangle}{\sqrt{2}}$ is
the singlet state and $F_{w}$ is the maximal singlet fraction
corresponding to the Werner state.
$F_{w}$ is also related to the linear entropy $S_{L}$ as
\begin{eqnarray}
F_{w}=\frac{1+3\sqrt{1-S_{L}}}{4}
\end{eqnarray}

The concurrence of $\rho_{W}$ is given by
\begin{eqnarray}
\textit{C}(\rho_{W})&=& max
\{0,2F_{w}-1\}\nonumber\\ &=&\left\{\begin{array}{cccc} 0
& & & 0\leq F_{w} \leq \frac{1}{2}\\
2F_{w}-1 & & & \frac{1}{2}< F_{w} \leq 1
\end{array}
\right. \label{werner-con.}
\end{eqnarray}

When the Werner state is used as a quantum channel for teleportation,
the average optimal teleportation fidelity is given by
\cite{horodecki1,badziag,mista}
\begin{eqnarray}
f_{opt}^{T}(\rho_{W})= \frac{2F_{w}+1}{3},~~~~~ \frac{1}{2}< F_{w}
\leq 1\label{werner.tel.fid.}
\end{eqnarray}
Similarly, the relation between the teleportation fidelity and
the concurrence of the Werner state is given by
\begin{eqnarray}
f_{opt}^{T}(\rho_{W})=\frac{2+C(\rho_{W})}{3}
\label{fidconc}
\end{eqnarray}
In terms of the linear entropy $S_{L}$, Eq.(\ref{werner.tel.fid.}) can be
re-written as
\begin{eqnarray}
f_{opt}^{T}(\rho_{W})= \frac{1+\sqrt{1-S_{L}}}{2},~~~~~ 0 \leq
S_{L} < \frac{8}{9}\label{werner.tel.fid.lin.entropy}
\end{eqnarray}
Further, we have
\begin{eqnarray}
F(\rho_{W})=\frac{1+N(\rho_{W})}{4} \label{fid12}
\end{eqnarray}
Now using the inequality (\ref{ineq123}) in equation
(\ref{fid12}), we have
\begin{eqnarray}
F(\rho_{W})\leq \frac{1}{2}[1+N^{W}]
\end{eqnarray}
which is the upper bound of the singlet
fraction for the Werner state in terms of negativity.

We now review the status of the violation of the
Bell-CHSH inequality by the Werner state.
Using Eq.(\ref{tmatrix}) the eigenvalues of the matrix
$T_{w}^{\dagger}T_{w}$  are given by
$u_{1}=u_{2}=u_{3}=\frac{(4F_{w}-1)^2}{9}$, where
$(T_{w})_{nm}=Tr(\rho_{W}\sigma_{n}\otimes\sigma_{m})$ denotes the
elements of the matrix $T_{w}$. The
Werner state violates the Bell-CHSH inequality iff
$M(\rho_{W})>1$, where $M(\rho_{W})$ is given by
\begin{eqnarray}
M(\rho_{W})=  2\frac{(4F_{w}-1)^2}{9}
\label{werner-m-rho}
\end{eqnarray}
Using Eq.(\ref{werner-con.}) it follows that the
Werner state satisfies the Bell-CHSH inequality although
it is entangled when the
maximal singlet fraction $F_{w}$ lies within the range
\begin{eqnarray}
\frac{1}{2}\le F_{w}\leq \frac{3+\sqrt{2}}{4\sqrt{2}}
\label{wernerBell-CHSH2}
\end{eqnarray}
The optimal teleportation fidelity in terms of $M(\rho)$ is given by
\begin{eqnarray}
f_{opt}^T(\rho_W) = \frac{\sqrt{\frac{M(\rho_W)}{2}} +1}{2}
\label{werner.fid.bell}
\end{eqnarray}
Moreover, from Eqs.(\ref{werner.tel.fid.}) and (\ref{werner.fid.bell})
it follows that the Werner state can be used as a quantum teleportation
channel (average optimal fidelity exceeding $2/3$) even without violating the
Bell-CHSH inequality in the above domain.

\section{Teleportation via the Munro-James-White-Kwiat maximally entangled mixed state}

Munro {\it et al.} \cite{munro,wei}
showed that there exist a class of states that have significantly
greater degree of entanglement for a given linear entropy than the
Werner state. In this section we will investigate whether the
class of states introduced by Munro {\it et al.} could be used as
a teleportation channel. We begin with the
analytical form of the MEMS given by
\begin{eqnarray}
\rho_{MEMS}= \left(\begin{matrix}{h(\textit{C}) & 0 & 0 &
\frac{\textit{C}}{2}\cr 0 & 1-2h(\textit{C})& 0 &0 \cr 0  & 0 &
0&0 \cr \frac{\textit{C}}{2} &0 & 0 & h(\textit{C})
}\end{matrix}\right) \label{T16}
\end{eqnarray}
where
\begin{eqnarray}
h(\textit{C})=\left\{\begin{array}{cccc}
\textit{C}/2 & & &\textit{C}\geq \frac{2}{3}\\
1/3 & & & \textit{C}< \frac{2}{3}
\end{array}
\right. \label{h(C)}
\end{eqnarray}
with $\textit{C}$ denoting the concurrence of $\rho_{MEMS}$ (\ref{T16}).\\
The form of the linear entropy is given by
\begin{eqnarray}
S_{L}=\left\{
\begin{array}{cccc}
\frac{8}{3}(\textit{C}-\textit{C}^{2}) & & &\textit{C}\geq 2/3\\
\frac{2}{3}(\frac{4}{3}-\textit{C}^{2}) & & & \textit{C}< 2/3
\end{array}
\right. \label{T17}
\end{eqnarray}

To see the performance of the MEMS state (\ref{T16}) as a
teleportation channel, we have to calculate the fidelity of the
teleportation channel. We use the
result given in Eq.(\ref{T18})
relating the optimal
teleportation fidelity  and the singlet fraction of a state $\rho$.
The maximal singlet fraction of the
state described by the density operator $\rho_{MEMS}$ using the
definition (\ref{sing.frac.})
is found out to be
\begin{eqnarray}
F_{MEMS}
&&=max\{h(\textit{C})+\frac{\textit{C}}{2},h(\textit{C})-\frac{\textit{C}}{2},\frac{1}{2}-h(\textit{C}),
\frac{1}{2}-h(\textit{C})  \nonumber \\
&&=h(\textit{C})+\frac{\textit{C}}{2}
\label{T20}
\end{eqnarray}
Using Eqs.(\ref{T18}) and (\ref{h(C)}), the optimal teleportation fidelity
is given by
\begin{eqnarray}
f_{opt}^{T}(\rho_{MEMS})=\left\{
\begin{array}{cccc}
\frac{2\textit{C}+1}{3} & & &\textit{C}\geq 2/3\\
\frac{5+3\textit{C}}{9} & & & \textit{C}< 2/3
\end{array}
\right. \label{Tel.fid.(C)}
\end{eqnarray}
Now inverting the relation (\ref{T17}), i.e., expressing
$\textit{C}$ in terms of $S_{L}$, we can rewrite
Eq.(\ref{Tel.fid.(C)}) in terms of the linear entropy $S_{L}$ as
\begin{eqnarray}
f_{opt}^{T}(\rho_{MEMS})=\left\{
\begin{array}{cccc}
\frac{2}{3}+\frac{\sqrt{2-3S_{L}}}{3\sqrt{2}} & & & 0\leq S_{L}\leq \frac{16}{27}\\
\frac{5}{9}+\frac{\sqrt{8-9S_{L}}}{3\sqrt{6}} & & & \frac{16}{27}<
S_{L} \leq \frac{8}{9}
\end{array}
\right. \label{Tel.fid.(S_{L})}
\end{eqnarray}
It follows that the MJKW \cite{munro} maximally entangled mixed state
(\ref{T16})
can be used as a faithful teleportation channel when the mixedness
of the state is less than the value $S_{L}=22/27$.

Note that since in specific cases of teleportation the teleportation fidelity 
depends upon the input states,  it
gives better results for some input states and worse for some other
input states. But here we  use the formula for average teleportation fidelity
averaging over all input states. However, for specific cases of input states 
it is possible to perform a calculation for the best (or the worst) 
teleportation fidelity (rather than the averaged optimum) as we illustrate now.
For example, if we 
consider the input state to be teleported  is of the form
\begin{eqnarray}
\rho^{in}= \left(\begin{matrix}{x & y \cr y^{*} &
1-x}\end{matrix}\right)
\end{eqnarray}
and if the teleportation channel is given by $\rho_{MEMS}$, 
the teleported state (using the standard teleportation
protocol) after performing suitable
unitary transformations corresponding to the four Bell-state
measurement outcomes $|\phi^{+}\rangle$,$|\phi^{-}\rangle$,
$|\psi^{+}\rangle$ and $|\psi^{-}\rangle$ is given by (for the following
two cases):\\
(i) For $C\geq\frac{2}{3}$
\begin{eqnarray}
\rho_{B_{1}}^{out}=\rho_{B_{2}}^{out}=
\left(\begin{matrix}{\frac{xC}{2N} & \frac{yC}{2N} \cr
\frac{y^{*}C}{2N} &
\frac{x(2-3C)+C}{2N}}\end{matrix}\right) \nonumber{}\\ \nonumber{}\\
\rho_{B_{3}}^{out}=-\rho_{B_{4}}^{out}=
\left(\begin{matrix}{\frac{(3x-2)C+2(1-x)}{2N_{1}} &
\frac{yC}{2N_{1}} \cr \frac{y^{*}C}{2N_{1}} &
\frac{(1-x)C}{2N_{1}}}\end{matrix}\right)
\end{eqnarray}
and,
(ii) for $C<\frac{2}{3}$
\begin{eqnarray}
\rho_{B'_{1}}^{out}=\rho_{B'_{2}}^{out}=
\left(\begin{matrix}{\frac{x}{3N} & \frac{yC}{2N} \cr
\frac{y^{*}C}{2N} &
\frac{1}{3N}}\end{matrix}\right) \nonumber{}\\ \nonumber{}\\
\rho_{B'_{3}}^{out}=-\rho_{B'_{4}}^{out}=
\left(\begin{matrix}{\frac{1}{3N_{1}} & \frac{yC}{2N_{1}} \cr
\frac{y^{*}C}{2N_{1}} & \frac{(1-x)}{3N_{1}}}\end{matrix}\right)
\end{eqnarray}
To determine the efficiency of the teleportation channel, we
calculate the distances between the input and output state using
Hilbert Schmidt norm, and they are given by \\
(i) for $C\geq\frac{2}{3}$
\begin{eqnarray}
D_{B_{1}}&=& D_{B_{2}} =
x^{2}(1-\frac{C}{2N})^{2}+2|y|^{2}(1-\frac{C}{2N})^{2} \nonumber\\
&+& [(1-x)-\frac{x(2-3C)+C}{2N}]^{2} \nonumber{}\\
D_{B_{3}} &=& x-\frac{(3x-2)C+2(1-x)}{2N_{1}}]^{2}+2|y|^{2}(1-\frac{C}{2N_{1}})^{2}
\nonumber\\ &+& (1-x)^{2}(1-\frac{C}{2N_{1}})^{2} \nonumber{}\\
D_{B_{4}} &=& x+\frac{(3x-2)C+2(1-x)}{2N_{1}}]^{2}+2|y|^{2}(1+\frac{C}{2N_{1}})^{2}\nonumber\\ &+&(1-x)^{2}(1+\frac{C}{2N_{1}})^{2}
\end{eqnarray}
where $N= x(1-\frac{C}{2})+\frac{(1-x)C}{2}$ and $N_{1}= \frac{xC}{2}+(1-x)(1-\frac{C}{2})$, and 
(ii) for $C<\frac{2}{3}$
\begin{eqnarray}
D_{B'_{1}}&=& D_{B'_{2}} = x^{2}(1-\frac{1}{3N'})^{2}+2|y|^{2}(1-\frac{C}{2N'})^{2}
\nonumber\\ &+& [(1-x)-\frac{1}{3N'}]^{2} \nonumber{}\\
D_{B'_{3}} &=& (x-\frac{1}{3N'_{1}})^{2}+2|y|^{2}(1-\frac{C}{2N'_{1}})^{2}
\nonumber\\ &+& (1-x)^{2}(1-\frac{1}{3N'_{1}})^{2} \nonumber{}\\
D_{B'_{4}} &=& (x+\frac{1}{3N'_{1}})^{2}+2|y|^{2}(1+\frac{C}{2N'_{1}})^{2}
\nonumber\\ &+&(1-x)^{2}(1+\frac{1}{3N'_{1}})^{2}
\end{eqnarray}
where where $N'= \frac{2x}{3}+\frac{1-x}{3}$ and $N'_{1}= \frac{x}{3}+\frac{2(1-x)}{3}$.
The teleportation fidelity $(F)$ can be easily calculated by using
the formula $F=1-D$. Clearly, the fidelity depends on the input
state and hence one can easily calculate the best (or worst)
fidelity by choosing some particular input state. However, the puprose
of the present paper is to compare the average perfomance of various
teleportation channels, and to this end henceforth in this work we will
deal further with average optimal teleportation fidelities only.

Next, we return to the nonlocal properties of the state $\rho_{MEMS}$.
Wei {\it et al.} \cite{wei} have studied the state $\rho_{MEMS}$ from
the perspective of Bell's-inequality violation. Here we focus on
the parametrization of the state given by Eq.(\ref{T16}) and
demarcate the range of concurrence where the  Bell-CHSH inequality is violated.
In order to use the result (\ref{bellineq}) we construct the matrix $T_{MEMS}$
as
\begin{eqnarray}
T_{MEMS}= \left(\begin{matrix}{h(\textit{C})+C & 0 & 0 \cr 0 & -C
& 0  \cr 0  & 0 & 4h(\textit{C})-1 }\end{matrix}\right)
\label{T100}
\end{eqnarray}
The eigenvalues of the matrix $(T_{MEMS}^{\dagger}T_{MEMS})$ are
given by
\begin{eqnarray}
u_{1}=(h(C)+C)^{2},~~ u_{2}=C^{2}, ~~
u_{3}=(4h(C)-1)^{2} \label{eigen}
\end{eqnarray}
In accord with Eq.(\ref{h(C)}), the eigenvalues (\ref{eigen})
take two different forms which are discussed separately below:\\
Case-I: $h(C)=\frac{C}{2},~~\frac{2}{3}\leq C \leq 1$.
The eigenvalues (\ref{eigen}) reduce to
\begin{eqnarray}
u_{1}=\frac{9C^{2}}{4},~~ u_{2}=C^{2},~~\textrm{and}~~~
u_{3}=(2C-1)^{2} \label{eigen1}
\end{eqnarray}
When $C\geq \frac{2}{3}$, the eigenvalues can be arranged as
$u_{1}>u_{2}>u_{3}$. Therefore,
\begin{eqnarray}
M(\rho_{MEMS})=u_{1}+u_{2}=\frac{13C^{2}}{4} \label{M}
\end{eqnarray}
One can easily see that $M(\rho_{MEMS})>1$
when $C \geq \frac{2}{3}$, and hence, in this case the state
$\rho_{MEMS}$ violates the Bell-CHSH inequality.\\
Case-II: $h(C)=\frac{1}{3},~~0\leq C <\frac{2}{3}$. The
eigenvalues given by Eq.(\ref{eigen}) reduce to
\begin{eqnarray}
u_{1}=\frac{(3C+1)^2}{9},~~ u_{2}=C^{2},~~\textrm{and}~~~
u_{3}=\frac{1}{9} \label{eigen2}
\end{eqnarray}
Now we can split the interval $0\leq C <\frac{2}{3}$ into two
sub-intervals $0\leq C \leq \frac{1}{3}$ and $\frac{1}{3} < C<\frac{2}{3}$,
where the ordering of the eigenvalues are different.\\
(i) when $0\leq C \leq \frac{1}{3}$, the ordering of the eigenvalues
are $u_{1}>u_{3}>u_{2}$. In this case one has
\begin{eqnarray}
M(\rho_{MEMS})-1=
u_{1}+u_{3}-1=\frac{9C^{2}+6C-7}{9} \label{M1}
\end{eqnarray}
From Eq.(\ref{M1}) it is clear that $M(\rho_{MEMS})<1$ when
$0\leq C \leq \frac{1}{3}$. Hence, the Bell-CHSH inequality is satisfied by
$\rho_{MEMS}$.\\
(ii) when $\frac{1}{3} < C<\frac{2}{3}$, the ordering of the
eigenvalues are $u_{1}>u_{2}>u_{3}$. Therefore, the expression for
($M(\rho_{MEMS})-1$) is given by
\begin{eqnarray}
M(\rho_{MEMS})-1=
u_{1}+u_{2}-1=\frac{2(9C^{2}+3C-4)}{9} \label{M2}
\end{eqnarray}
From Eq.(\ref{M2}), it follows that $M(\rho_{MEMS})>1$ when
$\frac{\sqrt{153}-3}{18}<C<\frac{2}{3}$
and hence the state $\rho_{MEMS}$ violates the Bell-CHSH inequality.
On the contrary, $M(\rho_{MEMS})\leq 1$ when
$\frac{1}{3}<C\leq \frac{\sqrt{153}-3}{18}$,
and hence the state $\rho_{MEMS}$
satisfies the Bell-CHSH inequality although it is entangled.
It was noticed earlier \cite{jmodopt} that the MJKW state needs a much higher 
degree of entanglement to violate the Bell-CHSH inequality compared to the 
Werner states. Our above results revalidate this fact.

We next consider a wider class of maximally entangled mixed states as
proposed by Wei et al. \cite{wei}. The general form of a two qubit density 
matrix
comprising a mixture of the maximally entangled Bell state
$|\Phi_{+}\rangle$ and a mixed diagonal state is given by
\begin{eqnarray}
\rho_{G}= \left(\begin{matrix}{x+\frac{\gamma}{2} & 0 & 0 &
\frac{\gamma}{2}\cr 0 & a & 0 & 0\cr 0 & 0 & b &
0\cr\frac{\gamma}{2}  & 0 & 0 &
y+\frac{\gamma}{2}}\end{matrix}\right)
\end{eqnarray}
where $a$, $b$, $x$, $y$ and $\gamma$ are non-negative real parameters. 
The normalization condition gives
$x+y+\gamma+a+b=1$
The entanglement of $\rho_{G}$ is quantified by
\begin{eqnarray}
C(\rho_{G})=max[\gamma-2\sqrt{ab},0]
\end{eqnarray}
Therefore, the state $\rho_{G}$ is entangled only if
$\gamma>2\sqrt{ab}$. The correlation matrix for  $\rho_{G}$ is
given by:
\begin{eqnarray}
T_{G}= \left(\begin{matrix}{\gamma & 0 & 0 \cr 0 & -\gamma & 0
\cr 0 & 0 & x+y+\gamma-a-b}\end{matrix}\right)
\end{eqnarray}
The eigen values of the symmetric matrix $T_{G}^{\dagger}T_{G}$ are
given by
$v_{1}=\gamma^{2}$, $v_{2}=\gamma^{2}$, and
$v_{3}=(x+y+\gamma-a-b)^{2}=(1-2a-2b)^{2}$. Now, the quantity
$M(\rho_{G})$ is given by
\begin{eqnarray}
M(\rho_{G})=max_{i>j}(v_{i}+v_{j})
\end{eqnarray}
Here one is led to the following two cases.
Case (i): $M(\rho_{G})=2\gamma^{2}$, when either $\gamma>2(a+b)-1$
and $\gamma>2\sqrt{ab}$, for $a+b>\frac{1}{2}$ or
$\gamma>1-2(a+b)$ and $\gamma>2\sqrt{ab}$ for $a+b<\frac{1}{2}$; and 
Case (ii): $M(\rho_{G})=\gamma^{2}+(1-2a-2b)^{2}$ when either
$2(a+b)-1<\gamma<2\sqrt{ab}$, for $a+b>\frac{1}{2}$ or
$2\sqrt{ab}<\gamma<1-2(a+b)$ for $a+b<\frac{1}{2}$.
In either case the Bell-CHSH inequality is violated  if $M(\rho_{G})>1$. 
 
Now, our task is to find the condition when the state
$\rho_{G}$ could be used as a teleportation channel. Hence, we
have to find the condition under which $N(\rho_{G})>1$. In this
case $N(\rho_{G})$ is given by
\begin{eqnarray}
N(\rho_{G})=\sqrt{v_{1}}+\sqrt{v_{2}}+\sqrt{v_{3}}=1+2(\gamma-a-b)
\label{nrhog}
\end{eqnarray}
Therefore, we have
\begin{eqnarray}
N(\rho_{G})>1\Rightarrow \gamma>a+b>2\sqrt{ab}
\end{eqnarray}
It follows from Eq.(\ref{nrhog}) that
\begin{eqnarray}
f_{opt}^{T}(\rho_{G})=\frac{1}{2}[1+\frac{N(\rho_{G})}{3}]=
\frac{2}{3}+\frac{1}{3}(\gamma-a-b)
\label{foptnrhog}
\end{eqnarray}
Writing the optimal teleportation fidelity in the above form enables a
useful comparison with the teleportation capacity of the Werner state.
Note that for either $a=0$, or $b=0$, one has $C(\rho_{W}) = \gamma$. Hence,
it follows that the average optimal teleportation fidelity of the Werner
state can be written as
\begin{eqnarray}
f_{opt}^T(\rho_W) = \frac{2}{3} + \frac{\gamma}{3}
\label{foptgammawer}
\end{eqnarray}
From Eqs.(\ref{foptnrhog}) and (\ref{foptgammawer}) it immediately follows
that 
\begin{eqnarray}
 f_{opt} ^{T}(\rho_{G})<f_{opt}^{T}(\rho_{werner})
\label{telfigwerrho}
\end{eqnarray}
which shows that the 
Werner state performs better as a teleportation channel than the
general MEMS.

\section{Non-maximally entangled mixed states as teleportation channels}

\subsection{The Werner Derivative}

Hiroshima and Ishizaka \cite{hiroshima} studied a particular class
 of mixed states -  Werner derivative - obtained by applying a
nonlocal unitary operator $U$ on the Werner state, i.e.,
$\rho_{wd} = U\rho_{W}U^{\dagger}$.
The Werner derivative is described by the
density operator
\begin{eqnarray}
\rho_{wd}=\frac{1-F_{w}}{3}I_{4}+\frac{4F_{w}-1}{3}|\psi\rangle\langle\psi|
\label{T8}
\end{eqnarray}
where
$|\psi\rangle=U|\Psi^{-}\rangle=\sqrt{a}|00\rangle+\sqrt{1-a}|11\rangle$
with $\frac{1}{2}\leq a\leq 1$. The state (\ref{T8}) is entangled
if and only if \cite{hiroshima}
\begin{eqnarray}
\frac{1}{2}\leq a <
\frac{1}{2}(1+\frac{\sqrt{3(4F_{w}^{2}-1)}}{4F_{w}-1}) \label{T8a}
\end{eqnarray}
which futher gives a restriction on $F_{w}$ as $\frac{1}{2}<F_{w} \leq 1$.

Our aim here is to study how efficiently the Werner derivative
works as a teleportation channel. To do this, let us start with
the matrix $T_{wd}$ for the state $\rho_{wd}$  given by
\begin{eqnarray}
T_{wd}=  \phantom{xxxxxxxxxxxxxxxxxxxxxxxxxxxxxxxxxx} \nonumber \\
 \left(\begin{matrix}{\frac{2\sqrt{a(1-a)}(4F_{w}-1)}{3}
& 0 & 0 \cr 0 & -\frac{2\sqrt{a(1-a)}(4F_{w}-1)}{3}& 0 \cr 0  & 0
& \frac{(4F_{w}-1)}{3}}\end{matrix}\right)\label{T9}
\end{eqnarray}
The eigenvalues of the matrix ($T^{\dagger}_{wd}T_{wd}$) are
$u_{1}=u_{2}=\frac{4a(1-a)(4F_{w}-1)^{2}}{9},u_{3}=\frac{(4F_{w}-1)^{2}}{9}$.
The  Werner Derivative can be used as a teleportation
channel if and only if it stisfies Eq.(\ref{teleportcond}), i.e.,
$N(\rho_{wd})>1$, where
\begin{eqnarray}
N(\rho_{wd})=\sqrt{u_{1}}+\sqrt{u_{2}}+\sqrt{u_{3}} \phantom{xxxxxxx} \nonumber \\
=\frac{(4F_{w}-1)[1+4\sqrt{a(1-a)]}}{3}
\label{T10}
\end{eqnarray}
It follows that the Werner Derivative can be used as a teleportation
channel if and only if
\begin{eqnarray}
16a^{2}-16a+\alpha^{2} < 0
\label{T11}
\end{eqnarray}
where $\alpha=\frac{4(1-F_{w})}{4F_{w}-1}$.
Solving (\ref{T11}) for the parameter $a$, we get
\begin{eqnarray}
\frac{1}{2}\leq a <\frac{1}{2}+\frac{\sqrt{4-\alpha^{2}}}{4}\equiv
\frac{1}{2}(1+\frac{\sqrt{3(4F_{w}^{2}-1)}}{4F_{w}-1})
\label{acond2}
\end{eqnarray}
Therefore, teleportation can be done faithfully via $\rho_{wd}$
when the parameter $\textit{a}$ satisfies the inequality
(\ref{T8a}).

The  fidelity of teleportation is given by
\begin{eqnarray}
f^{T}_{opt}(\rho_{wd}) && =\frac{1}{2}[1+\frac{1}{3}N(\rho_{wd})]\nonumber\\
&& =\frac{1}{18}[9+(4F_{w}-1)(1+4\sqrt{a(1-a)})] \phantom{xx}
\label{T13}
\end{eqnarray}
When $a=\frac{1}{2}$, the Werner derivative reduces to the Werner state,
and the teleportation fidelity  also
reduces to that of the Werner state given by Eq.(\ref{werner.tel.fid.}).
From Eq.(\ref{T13}), it is clear that $f^{T}_{opt}(\rho_{wd})$
is a decreasing function of $a$, and hence from Eq.(\ref{acond2}), one
obtains
\begin{eqnarray}
\frac{2}{3}<f^{T}_{opt}(\rho_{wd})\leq
\frac{2F_{w}+1}{3} \label{T14}
\end{eqnarray}
Further, we can express the teleportation fidelity
$f^{T}_{opt}(\rho_{wd})$ given in  Eq.(\ref{T13}) in terms of
linear entropy $S_{L}$ as
\begin{eqnarray}
f_{opt}^{T}(\rho_{wd}) \phantom{xxxxxxxxxxxxxxxxxxxxxxxxxxxxxxx} \nonumber \\
=\frac{9+3\sqrt{1-S_{L}}(1+4\sqrt{a(1-a)})}{18},~0\leq
S_{L}<\frac{8}{9} \label{T13c}
\end{eqnarray}

Now we investigate whether the state $\rho_{wd}$ violates the
Bell-CHSH inequality using the condition given in Eq.(\ref{bellineq}).
The real valued function $M(\rho)$ for the Werner derivative state
is given by
\begin{eqnarray}
M(\rho_{wd})=u_{2}+u_{3}=\frac{(1+4a-4a^2)(4F_{w}-1)^{2}}{9}
\label{M(Wer.der.)}
\end{eqnarray}
It follows that
\begin{eqnarray}
M(\rho_{wd})-1 = \frac{-(4F_{w}-1)^2}{9} (a-\beta)(a-\gamma)\label{M(Bell)}
\end{eqnarray}
where
\begin{eqnarray}
\beta && =\frac{1}{2}(1-\frac{\sqrt{2(4F_{w}-1)^2-9}}{4F_{w}-1})\nonumber \\
\gamma && =\frac{1}{2}(1+\frac{\sqrt{2(4F_{w}-1)^2-9}}{4F_{w}-1})
\label{beta,gamma}
\end{eqnarray}
For $\beta$ and $\gamma$ to be real,
$\frac{3+\sqrt{2}}{4\sqrt{2}}\leq F_{w}\leq 1$.
From the expression of $\beta$ and Eq.(\ref{T8a}), it is clear that $\beta\leq
\frac{1}{2}\leq
\textit{a}<\frac{1}{2}(1+\frac{\sqrt{3(4F_{w}^{2}-1)}}
{4F_{w}-1})$ as $\frac{3+\sqrt{2}}{4\sqrt{2}} \leq F_{w}\leq1$. Hence $\textit{a}-\beta\geq0$.
Next, from the expression of $\gamma$, it follows that
$\gamma \leq \frac{1}{2}(1+\frac{\sqrt{3(4F_{w}^{2}-1)}} {4F_{w}-1})$.
Now, we consider the following three cases separately:\\
Case-I: If
$\gamma<\textit{a}<\frac{1}{2}(1+\frac{\sqrt{3(4F_{w}^{2}-1)}}
{4F_{w}-1})$ and $\frac{3+\sqrt{2}}{4\sqrt{2}}< F_{w}\leq1$, then
$M(\rho_{wd})-1<0$. In this case the
Bell-CHSH inequality is respected by the state
$\rho_{wd}$ although the state is entangled there.\\
Case-II: If $\frac{1}{2}\leq \textit{a}<\gamma$ and
$\frac{3+\sqrt{2}}{4\sqrt{2}}< F_{w}\leq1$, then $M(\rho_{wd})-1>0$.
Thus in this range of the parameter
$\textit{a}$ the Bell-CHSH inequality is violated by
the state $\rho_{wd}$.\\
Case-III: Here we consider the situation when $F_{w}=
\frac{3+\sqrt{2}}{4\sqrt{2}}$. In this case
$\beta=\gamma=\frac{1}{2}$ and hence $M(\rho_{wd})\leq 1$ holds
for $\frac{1}{2}\leq a < \frac{1}{2}(1+\frac{\sqrt{1+2\sqrt{2}}}
{2}))$. The equality sign is achieved when
$a=\beta=\gamma=\frac{1}{2}$. Therefore, in the case when $F_{w}=
\frac{3+\sqrt{2}}{4\sqrt{2}}$ the Werner derivative satisfies the
Bell-CHSH inequality although it is entangled.

\subsection{A new non-maximally entangled mixed state}

We construct a two-qubit density matrix
$\rho_{new}$ as a convex combination of a separable density matrix
$\rho^{G}_{12}=Tr_{3}(|GHZ\rangle_{123})$ and an inseparable
density matrix $\rho^{W}_{12}=Tr_{3}(|W\rangle_{123})$ where
$|GHZ\rangle$ and $|W\rangle$ denote the three-qubit GHZ-state\cite{ghz}
and the W-state\cite{wstate} respectively. This construction is somewhat
similar in spirit to the Werner state which is a convex combination of
a maximally mixed state and a maximally entangled pure state. 
We exploit here the
properties that the GHZ state and the W state are two qubit separable and
inseparable states, respectively, when a qubit is lost from the corresponding
three qubit states. By
constructing this type of a non-maximally entangled mixed state, 
our aim is to show that it can be used as a better teleportation channel 
compared to the
Werner derivative state.

The two-qubit state described by the density matrix $\rho_{new}$
can be explicitly written as
\begin{eqnarray}
\rho_{new}=p\rho^{G}_{12}+(1-p)\rho^{W}_{12},~~~~0\leq p \leq 1
\label{T1}
\end{eqnarray}
The matrix representation of the density matrix $\rho_{new}$ in
the computational basis is given by
\begin{eqnarray}
\rho_{new} = \left(\begin{matrix}{\frac{p+2}{6} & 0 & 0 & 0 \cr 0
& \frac{1-p}{3} & \frac{1-p}{3} & 0 \cr 0 & \frac{1-p}{3}&
\frac{1-p}{3} & 0 \cr 0 & 0 & 0 & \frac{p}{2}
}\end{matrix}\right). \label{T2}
\end{eqnarray}
Since the state described by the density matrix (\ref{T2}) is of
the form
\begin{eqnarray}
\sigma = \left(\begin{matrix}{a & 0 & 0 & 0 \cr 0 & b & c & 0 \cr
0 & c^{*}& d & 0 \cr 0 & 0 & 0 & e }\end{matrix}\right) \label{T3}
\end{eqnarray}
its amount of entanglement \cite{bruss} is given by
\begin{eqnarray}
C(\rho_{new}) && =C(\sigma)=2max(|c|-\sqrt{ae},0) \nonumber\\
&& = 2max((\frac{1-p}{3}-\sqrt{\frac{p(p+2)}{12}}),0) \label{T4}
\end{eqnarray}
Therefore, $\rho_{new}$ is entangled only if
$\frac{1-p}{3}-\sqrt{\frac{p(p+2)}{12}}>0$, i.e., when
$0\leq p<0.292$.

Note that in the limiting case of
$p=0$ the state $\rho_{new}$ reduces to
\begin{eqnarray}
\rho_{12}^{W} =
\frac{1}{3}|00\rangle\langle00|+\frac{2}{3}|\psi^{+}\rangle\langle\psi^{+}|
\label{mat.p=0}
\end{eqnarray}
where $|\psi^+\rangle = (|01\rangle + |10\rangle)/\sqrt{2}$.
The state
$\rho_{12}^{W}$ is maximally entangled since it can be put into Ishizaka and
Hiroshima's \cite{ishizaka} proposed
class of MEMS. The concurrence of this state is $\frac{2}{3}$. When
this state is used as a teleportation channel,  the
teleportation fidelity becomes
$f^{T}_{opt}(\rho_{12}^{W})=\frac{7}{9}$. Moreover, it can be
checked that the state
$\rho_{12}^{W}$ satisfies the Bell-CHSH inequality although it is an
entangled state.

To obtain the teleportation fidelity for the
state $\rho_{new}$, we first construct the matrix $T_{new}$ using
Eq.(\ref{tmatrix}), which
is given by
\begin{eqnarray}
T_{new} = \left(\begin{matrix}{\frac{2(1-p)}{3} & 0 & 0 \cr 0 &
\frac{2(1-p)}{3} & 0 \cr 0 & 0 & \frac{(4p-1)}{3}
}\end{matrix}\right) \label{T5}
\end{eqnarray}
The eigenvalues of ($T^{\dagger}_{new}T_{new}$) are given by
$u_{1}=u_{2}=\frac{4(1-p)^{2}}{9}$ and
$u_{3}=\frac{(4p-1)^{2}}{9}$.
When $p>\frac{1}{4}$, one has
$N(\rho_{new})=\sqrt{u_{1}}+\sqrt{u_{2}}+\sqrt{u_{3}}=1$.
Therefore, the teleportation fidelity becomes
$f^{T}_{opt}(\rho_{new})=\frac{1}{2}[1+\frac{1}{3}N(\rho_{new})]=\frac{2}{3}$.
Hence for $p>\frac{1}{4}$, the state $\rho_{new}$ cannot be used
as an efficient teleportation channel since it does
not overtake the classical fidelity. But when $0\leq
p<\frac{1}{4}$, $N(\rho_{new})=\frac{5-8p}{3}>1$, and hence
$\rho_{new}$ can be used as an efficient teleportation channel. In
this case the average optimal teleportation fidelity is given by
\begin{eqnarray}
f^{T}_{opt}(\rho_{new})= \frac{7-4p}{9},~~~~0\leq p <\frac{1}{4}
 \label{T6}
\end{eqnarray}
and it follows that
\begin{eqnarray}
\frac{2}{3}<f^{T}_{opt}(\rho_{new})\leq \frac{7}{9}
 \label{T7(a)}
\end{eqnarray}
We note here an interesting fact that the state $\rho_{new}$
cannot be used as an efficient teleportation channel when
$0.25<p<0.292$ although the state is entangled there.

When $\rho_{new}$ is used as a quantum teleportation channel
the mixedness of the state is given by
\begin{eqnarray}
S_{L}= \frac{2}{27}(8+14p-13p^{2}),~~~ 0\leq p <
\frac{1}{4}\label{li.ent.(new1)}
\end{eqnarray}
Therefore, the teleportation fidelity $f^{T}_{opt}(\rho_{new})$ in
terms of $S_{L}$ is given by
\begin{eqnarray}
f_{opt}^{T}(\rho_{new})=\frac{7-\frac{4}{26}(14-\sqrt{612-702S_{L}})}{9}, \nonumber \\
\frac{208}{351}\leq
S_{L} < \frac{2223}{2808} \label{T13b}
\end{eqnarray}

Let us now address  the
question as to whether the state $\rho_{new}$ violates the Bell-CHSH
inequality. We again
calculate the real valued function $M(\rho_{new})$ for the state
$\rho_{new}$ for the two following cases separately.\\
Case-I: When $0\leq p<\frac{1}{2}$,
$M(\rho_{new})=u_{1}+u_{2}=\frac{8+8p^{2}-16p}{9}$. Substituting
the values of $p$ in the above range it is easy to see that
$M(\rho_{new})\leq 1$, i.e., the Bell-CHSH inequality is satisfied.\\
Case-II: When $\frac{1}{2}\leq p \leq 1$,
$M(\rho_{new})=u_{1}+u_{3}=\frac{20p^{2}-16p+5}{9}$. It easily follows
that for the given range of values of $p$,
one has $M(\rho_{new}) \leq 1$.\\
Therefore, we conclude that in any case (i.e. $0\leq p \leq 1$),
the constructed state $\rho_{new}$ does not violate the Bell-CHSH
inequality although it is entangled for $0\leq p<0.292$.

\section{Comparison of teleportation fidelities for different
mixed states}

In the earlier sections we have studied the teleportation
capacities of various maximally as well as non-maximally entangled
mixed channels. It would be interesting now to actually compare
their performance in terms of the average optimal fidelities
corresponding to their respective magnitudes of entanglement, mixedness,
and also in relation
to their nonlocality properties manifested by the violations of
the Bell-CHSH inequality. Let us first compare the two MEMS states, viz.,
the Werner state and the MJKW state, whose average optimal teleportation
fidelities in terms of their respective concurrences are
given by (Eqs.(\ref{fidconc}) and (\ref{Tel.fid.(C)})),
\begin{eqnarray}
f_{opt}^{T}(\rho_{W})=\frac{2+C(\rho_{W})}{3} \\
f_{opt}^{T}(\rho_{MEMS})=\left\{
\begin{array}{cccc}
\frac{2\textit{C}+1}{3} & & &\textit{C}\geq 2/3\\
\frac{5+3\textit{C}}{9} & & & \textit{C}< 2/3
\end{array}
\right. \nonumber
\label{comp.fid.ent.}
\end{eqnarray}
In Fig.1 we plot $f_{opt}^T$ versus $C$ respectively for these two
MEMS states. One can see that the Werner state performs better as
a teleportation channel compared to the MJKW state for any given amount
of entanglement. Note further, that the MJKW state is useful for
teleportation ($f_{opt}^T(\rho_{MEMS}) > 2/3$) only when $C > 1/3$, whereas
the Werner state is able to serve as a quantum teleportation channel for
any amount of its entanglement.

\begin{figure}[h!]
\begin{center}
\includegraphics[width=11cm]{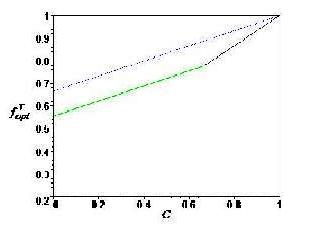}
\caption{(Coloronline) The average optimal teleportation fidelities
for the channels $\rho_W$ (dotted line) and $\rho_{MEMS}$ (full line) are
plotted with respect to
their respective magnitudes of entanglement $C$. Note that $\rho_{MEMS}$
performs as a quantum channel only for $C > 1/3$.}
\end{center}
\label{f1}
\end{figure}

Next, we compare the efficiency of teleportation of the two MEMS states
with respect to their nonlocality properties. The average teleportation
fidelities corresponding to the Werner state and the MJKW state are plotted
versus the function $M(\rho)$ in Fig.2. Since $M(\rho) > 1$ signifies the
violation of the Bell-CHSH inequality, it can be seen from the figure that
the MJKW state violates the Bell-CHSH inequality and simultaneously performs
as a teleportation channel in a certain region of parameter space, as
outlined in Section III. This is in contrast to the behaviour of the
Werner state which satisfies the Bell-CHSH inequality (in the parameter space
obtained in Section II) but yet performs as a quantum teleportation channel.
Next, in Fig.3 we present a comparison of the Werner state and the generalized
MEMS state $\rho_G$ by plotting respecively their average teleportation
fidelities versus the function $M(\rho)$.  We observe that the Werner State
and the general MEMS state are both useful for teleportation whether
they violate the Bell-CHSH inequality or not. But, the Werner state always 
performs better as
as a teleportation channel compared to the general MEMS $\rho_{G}$, except
at the value of $M(\rho)=1.7672$ where the
teleportation fidelities for both the states are the same. The result that
Werner states perform better as teleportation channels compared to the
other MEMS class of states can also be understood in terms of their
respective negativities $N$, i.e., $N^{MJWK}<N^{W}$ \cite{wei}.

\begin{figure}[h!]
\begin{center}
\includegraphics[width=11cm]{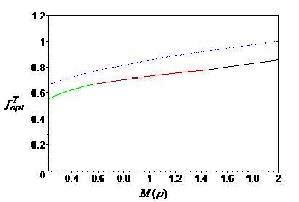}
\caption{(Coloronline) The average optimal teleportation fidelities
for the channels $\rho_W$ (dotted line) and $\rho_{MEMS}$ (full line) are
plotted with respect to
the quantity $M(\rho)$ indicating the nonlocal property of the channel.
$M(\rho) > 1$ signifies the violation of the Bell-CHSH inequality.}
\end{center}
\label{f2}
\end{figure}

The relationship between the mixedness of a channel and its ability to
perform quantum teleportation is one of the focal points of investigation
in this paper.
For this purpose, let us recall
the expressions (\ref{werner.tel.fid.lin.entropy}),
(\ref{Tel.fid.(S_{L})}), (\ref{T13c}) and (\ref{T13b}) for the
teleportation fidelities in terms of the linear entropy for all the
four types of states studied by us:
\begin{eqnarray}
&&f_{opt}^{T}(\rho_{new})=\frac{7-\frac{4}{26}(14-\sqrt{612-702S_{L}})}{9}, \nonumber\\
&& \frac{208}{351}\leq
S_{L} < \frac{2223}{2808} {}\nonumber\\&&
f_{opt}^{T}(\rho_{wd})=\frac{9+3\sqrt{1-S_{L}}(1+4\sqrt{a(1-a)})}{18}, \nonumber\\
&& 0\leq
S_{L}<\frac{8}{9}{} \nonumber\\&& f_{opt}^{T}(\rho_{MEMS})=\left\{
\begin{array}{cccc}
\frac{2}{3}+\frac{\sqrt{2-3S_{L}}}{3\sqrt{2}} & & & 0\leq S_{L}\leq \frac{16}{27}\\
\frac{5}{9}+\frac{\sqrt{8-9S_{L}}}{3\sqrt{6}} & & &
\frac{16}{27}\leq S_{L} \leq \frac{8}{9}
\end{array}
\right. {}\nonumber\\&& f_{opt}^{T}(\rho_{W})=
\frac{1+\sqrt{1-S_{L}}}{2},~~~~~ 0 \leq S_{L} < \frac{8}{9}
\label{comp.tel.fid.}
\end{eqnarray}

\begin{figure}[h!]
\begin{center}
\includegraphics[width=11cm]{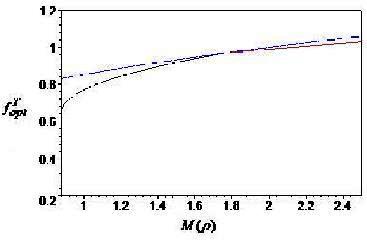}
\caption{(Coloronline) The average optimal teleportation fidelities
for the channels $\rho_W$ (dotted line) and $\rho_{G}$ (full line) are
plotted with respect to
the quantity $M(\rho)$ indicating the nonlocal property of the channel.
$M(\rho) > 1$ signifies the violation of the Bell-CHSH inequality.}
\end{center}
\label{f3}
\end{figure}

We first consider the comparison between the two maximally entangled
states, viz., the Werner state\cite{werner} $\rho_W$ and the MJKW state
\cite{munro} $\rho_{MEMS}$. From the above expressions of $f_{opt}^{T}$ for
these two states it follows
that $f_{opt}^{T}(\rho_{W}) = f_{opt}^{T}(\rho_{MEMS})$
only for $S_L = 0$. For all finite degrees of mixedness,
$f_{opt}^{T}(\rho_{W}) > f_{opt}^{T}(\rho_{MEMS})$. The two respective fidelities
are plotted versus the linear entropy in Fig.4. The MJKW state can be used
as a quantum teleportation channel only when its mixedness is less than
$S_L < 22/27$.  Although both these states
could perform as quantum teleportation channels for a range of values
of mixedness, one sees that the Werner state outperforms the MJKW state
for all finite values of mixedness even though the latter is more
entangled for specific values of linear entropy\cite{wei}. This is an
interesting result showing that all the entanglement of the MJKW class
of states is less useful as a resource for teleportation.

\begin{figure}[h!]
\begin{center}
\includegraphics[width=11cm]{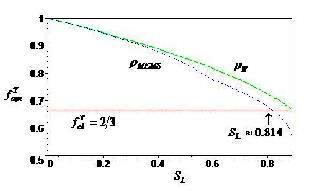}
\caption{(Coloronline) The average optimal teleportation fidelities
for the channels $\rho_W$ and $\rho_{MEMS}$ are plotted with respect to
the linear entropy $S_L$. The horzintal line represents the maximum
classical fidelity.}
\end{center}
\label{f4}
\end{figure}

Let us now devote our attention to the comparison of the two non-maximally
entangled mixed states $\rho_{wd}$ and $\rho_{new}$ that we have
studied in this paper. To address the issue as to
which of $\rho_{wd}$ and $\rho_{new}$ is more efficient as resource for
teleportation, we derive ranges for the parameters for which the
condition $N(\rho_{new})> N(\rho_{wd})$ holds such that the
teleportation fidelity via the channel $\rho_{new}$ will be greater than the
teleportation fidelity via $\rho_{wd}$. Here we make use of the
relationship between the teleportation fidelity
and the quantity $N(\rho)$ \cite{horodecki} given by Eq.(\ref{tel.fid.N}).
In the previous sections we have calculated $N(\rho_{wd})$ and
$N(\rho_{new})$, and their expressions are given by
\begin{eqnarray}
N(\rho_{wd})=\frac{(4F_{w}-1)(1+4\sqrt{a(1-a)})}{3},~~
\frac{1}{2}<F_{w}\leq 1 \label{N(wer.der.)}
\end{eqnarray}
\begin{eqnarray}
N(\rho_{new})=\frac{5-8p}{3},~~~~0\leq p<\frac{1}{4}\label{N(new)}
\end{eqnarray}
where the parameter $a$ lies within the range specified in Eq.(\ref{T8a}).
The state $\rho_{new}$ performs
better as a quantum channel for teleportation compared to
the state $\rho_{wd}$ only when $N(\rho_{new})> N(\rho_{wd})$,
from which using Eqs.(\ref{N(wer.der.)}) and (\ref{N(new)}) it follows that
\begin{eqnarray}
p<1-(\frac{1+2F_{w}}{4}+\frac{(4F_{w}-1)\sqrt{a(1-a)}}{2})
\label{c1}
\end{eqnarray}
One can easily verify that the condition (\ref{c1}) on the value of
$p$ is compatible with the upper bound on $p$ in Eq.(\ref{N(new)}).
However, consistency with the lower bound ($p>0$) imposes the following
conditions on the parameters $\textit{F}_{w}$ and $\textit{a}$:
\begin{eqnarray}
\frac{1}{2}+\frac{\sqrt{(F_{w}+1)(3F_{w}-2)}}{4F_{w}-1} < a &&<
\frac{1}{2}(1+\frac{\sqrt{3(4F_{w}^{2}-1)}}{4F_{w}-1}), \nonumber\\
&& F_w>\frac{2}{3} \label{c2}
\end{eqnarray}
Therefore, when the parameters $\textit{F}_{w}$, $\textit{a}$ and
$\textit{p}$ satisfy the relations given in  Eqs.(\ref{c1})
and (\ref{c2}),  one has $f^T_{opt}(\rho_{new}) > f^T_{opt}(\rho_{wd})$.

In order to understand better the comparitive performance of the two NMEMS
channels, let us rephrase our above arguments. For $\rho_{new}$ to perform 
better than $\rho_{wd}$,
we must have
$f_{opt}^{T}(\rho_{new})>f_{opt}^{T}(\rho_{wd})$, which in turn
implies that
$N(\rho_{new})>N(\rho_{wd})$ since $N(\rho)$ is related to the teleportation
fidelity of a channel by
$f_{opt}^{T}(\rho)=\frac{1}{2}[1+\frac{N(\rho)}{3}])$.
We have shown that the relation $f_{opt}^{T}(\rho_{new})>f_{opt}^{T}(\rho_{wd})$
holds true only when the inequality (\ref{c1}) is satisfied with
appropriate choices of  $a$ and $F_{w}$ which are
the parameters of the state $\rho_{wd}$, and $p$ which
is a parameter of our constructed state $\rho_{new}$
depends. The message that one obtains from the above calculations is that
similar to the case of maximally entangled mixed states (Werner and MJKW),
one can also construct different classes of non-maximally entangled mixed
states where one class can outperform another as a teleportation channel
depending upon the chosen parameters of the states.

Next we consider the situation in which the
$\rho_{wd}$ violates the Bell-CHSH inequality but $\rho_{new}$ satisfies it.
In this case let us see if
the teleportation fidelity
$f^{T}_{opt}(\rho_{wd})$ could still be less than the teleportation fidelity
$f^{T}_{opt}(\rho_{new})$. We have earlier shown in Section IV that $\rho_{new}$
satisfies the Bell-CHSH inequality, and we have also derived the ranges
for the parameters $a$ and $F_w$ for which $\rho_{wd}$ violates the
inequality. Combining these conditions with the requirements (\ref{c1})
and (\ref{c2}), we obtain several possible values for the parameters
$a$, $F_w$ and $p$ for which $f^T_{opt}(\rho_{new}) > f^T_{opt}(\rho_{wd})$.
These are listed in Table-I.

\begin{table}
\begin{tabular}{| c| c| c| c| c|}
\hline
  $\textit{F}_{w}$ & $\textit{a}$ & $\textit{p}$ & $f^{T}_{opt}(\rho_{wd})$
   & $f^{T}_{opt}(\rho_{new})$\\
    \hline
  0.96 & 0.962437 & 0.000006 & 0.777775 & 0.777775 \\
  \hline
   & 0.962490 & 0.000189 & 0.777694 & 0.777694 \\
  \hline
   & 0.970142 & 0.028321 & 0.765190 & 0.765191 \\
  \hline
   & 0.970144 & 0.028320 & 0.765187  & 0.765191\\
   \hline
 &  &  &  &  \\
  \hline
  0.97 & 0.964903 & 0.000003 & 0.777776 & 0.777776\\
  \hline
   & 0.964990 & 0.000320 & 0.777635 & 0.777636\\
  \hline
   & 0.978256  & 0.054980 & 0.753341 & 0.753342\\
  \hline
   & 0.978258 & 0.054980  & 0.753338 & 0.753342\\
       \hline
  &  &  &  &  \\
  \hline
  0.98 &0.967213 & 0.000004 & 0.777775 & 0.777776 \\
  \hline
   & 0.967290 & 0.000290 & 0.777644 & 0.777649\\
  \hline
   & 0.985910 & 0.087920 & 0.738701 & 0.738702\\
  \hline
   &0.985913 &  0.087938 & 0.738693 & 0.738694\\
       \hline
  &  &  &  &  \\
  \hline
  0.99 &0.969377 & 0.000004 &0.777776 & 0.777776 \\
  \hline
   & 0.969390 & 0.000056 & 0.777752 & 0.777753\\
  \hline
   & 0.993147 & 0.132901 &  0.718710 & 0.718711\\
  \hline
   & 0.993149 & 0.13291 &  0.718703 & 0.718707\\
       \hline
\end{tabular}
\caption{Comparison of teleportation fidelities
when $\rho_{wd}$ violates the Bell-CHSH inequality while $\rho_{new}$
satisfies it.}
\end{table}

We now present together the comparitive performance of
all the four entangled mixed states that we have considered in this
paper. We obtain the average optimal teleportation fidelities of
$\rho_W$, $\rho_{MEMS}$, $\rho_{wd}$ and $\rho_{new}$ in terms of
their linear entropies. Here we clearly address the question as
to how they compete as teleportation resources for specified
values of mixedness. The expressions for the teleportation fidelities
of $\rho_W$, $\rho_{MEMS}$ and $\rho_{new}$ are provided
explicitly in terms of the linear
entropy $S_L$ in Eqs.({\ref{comp.tel.fid.}). But for the state $\rho_{wd}$
we first obtain $F_{w}$ for a given $S_L$ using the relation
$F_{w} = \frac{1+3\sqrt{1-S_{L}}}{4}$. We then select a couple of values for
the parameter $a$ which lies in the range given in Eq.(\ref{c2}).
Since the mixedness of the state $\rho_{wd}$ does not depend on
the parameter $\textit{a}$, there exists a family of states
$\rho_{wd}$ for a given mixedness.
Finally the corresponding value of $f^T_{opt}(\rho_{wd})$ is
computed using the relation provided in Eq.(\ref{comp.tel.fid.}).
Our results are presented in Table-II.

\begin{table}
\begin{tabular}{| c| c| c| c| c| c| }
\hline
  $S_{L}$ & $\textit{a}$ &  $f^{T}_{opt}(\rho_{W})$&
  $f^{T}_{opt}(\rho_{MEMS})$& $f^{T}_{opt}(\rho_{wd})$ &$f^{T}_{opt}(\rho_{new})$\\
   \hline
  0.593 & 0.82  & 0.818983 & 0.777625 & 0.769726 & 0.777603 \\
  \hline
  0.593 & 0.95  &  0.818983 & 0.777625 & 0.6990218 & 0.777603 \\
   \hline
  0.600 &  0.80  &  0.816228 & 0.774982 & 0.774064 & 0.774581\\
   \hline
  0.600 &  0.93  &  0.816228 & 0.774982 & 0.712989 & 0.774581\\
  \hline
  0.62 & 0.77  & 0.808221 & 0.767251 & 0.775686 & 0.765728 \\
   \hline
  0.62 & 0.90  & 0.808221 & 0.767251 & 0.726029 & 0.765728 \\
  \hline
  0.64 & 0.74 & 0.800000 & 0.759226 & 0.775454 & 0.756516\\
   \hline
  0.64 & 0.85 & 0.800000 & 0.759226 & 0.742823 & 0.756516\\
  \hline
  0.66 & 0.70  & 0.791548 & 0.750871 & 0.775321 & 0.746896\\
    \hline
  0.66 & 0.92  & 0.791548 & 0.750871 & 0.702642 & 0.746896\\
   \hline
   \end{tabular}
\caption{Comparison of teleportation fidelities
for different MEMS and NMEMS channels for a given mixedness.}
\end{table}

The values for the linear entropy for which the
corresponding teleportation fidelities are displayed in Table-II are chosen
such that all the salient features of our results that we wish to
highlight are revealed in the range chosen. As expected, the MEMS states
perform better as teleportation channels in general compared to the NMEMS
states, with the Werner state giving rise to higher teleportation fidelity
for all values of mixedness. The comparison between the two NMEMS
states is affected by the fact that for a given $S_L$ there exists a family
of states $\rho_{wd}$ corresponding to different admissible values of
$a$. In this range some $\rho_{wd}$ states perform better compared
to $\rho_{new}$, but the situation may be reversed for a different
value of $a$ corresponding to the same value of mixedness. Moreover,
since mixedness for $\rho_{wd}$ does not depend on $a$, there exist
some values of $a$ for which $\rho_{wd}$ even outperforms the MJKW
state $\rho_{MEMS}$, as displayed in the Table.

\section{Conclusions}

To summarize, in this paper we have studied the efficiency of
maximally (MEMS) and non-maximally (NMEMS) entangled mixed states as
resources for teleportation. Since not every mixed entangled state is
useful for teleportation \cite{horodecki}, we have addressed
here the following questions.
Is every maximally entangled mixed state useful for teleportation ?
We answer this question in the negative by providing the example of
the maximally entangled MJKW \cite{munro} class of states which is not
useful for teleportation when its mixedness exceeds a certain bound.
Another question that we have investigated here is the relation between
the amount of entanglement for a state and its efficiency as a teleportation
channel. Our results show that a state which is
less entangled for a given degree of mixedness, e.g., the Werner
state \cite{werner}, could act as a more
efficient teleportation channel compared to a state that is more entangled,
e.g., the MJKW state.

One of our motivations here has been to compare the
performance of mixed entangled states as teleportation channels for
a specified amount of mixedness on one hand, and the amount of entanglement
on the other. We have
considered two specific
well-known MEMS, viz., the Werner state
and the MJWK  state, and obtained their average
teleportation fidelities in terms of their respective concurrences,
and also in terms of their respective linear entropies. In spite of the
fact that both these states fall in the category of maximally entangled
mixed states, we find that one of the them, viz., the Werner state, outperforms
the other, viz., the Munro state for either any fixed degree of mixedness,
or any specified magnitude of entanglement.
We have further considered two more class of
mixed states that are not maximally entangled (NMEMS). We have shown
that the Werner derivative \cite{hiroshima} can act as an efficient
quantum teleportation channel (with its average teleportation fidelity
exceeding the classical bound of $2/3$) in certain ranges of parameter
values. We then ask the question as
to whether there exist other class of NMEMS that could outperform
the Werner derivative as a teleportation resource. We answer this
question in the affirmative by constructing a new
non-maximally entangled mixed state
which is a convex combination of a separable state and an entangled state.

We have further investigated the issue as to whether the nonlocal properites
of quantum states, as characterized by the violation of local inequalities,
have any bearing on the
ability of mixed states to teleport efficiently \cite{popescu}. For MEMS
states, our analysis shows that the Werner state satisfies
the Bell-CHSH inequality and yet performs as a quantum teleportation
channel in a certain range of parameter space. We then ask the specific
question: is there any NMEMS state which does not violate the Bell-CHSH
inequality but is still useful for teleportation ? In this context we
first derive the conditions on the parameters for
which the Werner derivative state \cite{hiroshima}
satisfies the criterion of nonlocality
by violating the Bell-CHSH inequality. We then show that our constructed
new NMEMS state could perform as a quantum teleportation channel in spite
of satisfying the Bell-CHSH inequality. Moreover, our constructed state
yields a higher teleportation
fidelity compared to the Werner derivative even for a range of parameter
values where the latter violates the Bell-CHSH inequality.
We conclude by noting that for both maximally and non-maximally entangled
mixed states neither the magnitude of entanglement nor the violation of
local inequalities may be good indicators of their ability
to perform quantum information processing tasks such as teleportation.

\end{document}